\documentclass[sigconf]{acmart}

\usepackage{booktabs} 
\usepackage{todonotes}

\setcopyright{rightsretained}

\acmConference[]{}{June 2017}{}

\begin{document}
\title{Recurrent Latent Variable Networks for Session-Based Recommendation}

\author{Sotirios P. Chatzis}
\affiliation{%
 \institution{Department of Electrical Eng., Computer Eng., and Informatics}
 \institution{Cyprus University of Technology}
  \streetaddress{33 Saripolou Str.}
  \city{Limassol} 
  \state{Cyprus} 
  \postcode{3036}
}
\email{sotirios.chatzis@cut.ac.cy}

\author{Panayiotis Christodoulou}
\affiliation{%
  \institution{Department of Electrical Eng., Computer Eng., and Informatics}
  \institution{Cyprus University of Technology}
  \streetaddress{33 Saripolou Str.}
  \city{Limassol} 
  \state{Cyprus} 
  \postcode{3036}
}
\email{paa.christodoulou@edu.cut.ac.cy}

\author{Andreas Andreou}
\affiliation{%
  \institution{Department of Electrical Eng., Computer Eng., and Informatics}
  \institution{Cyprus University of Technology}
  \streetaddress{33 Saripolou Str.}
  \city{Limassol} 
  \state{Cyprus} 
  \postcode{3036}
}
\email{andreas.andreou@cut.ac.cy}

\begin{abstract}
In this work, we attempt to ameliorate the impact
of data sparsity in the context of session-based recommendation.
Specifically, we seek to devise a machine learning mechanism capable 
of extracting subtle and complex underlying temporal dynamics in the 
observed session data, so as to inform the recommendation algorithm.
To this end, we improve upon systems that utilize deep learning techniques with recurrently 
connected units; we do so by adopting concepts from the field of Bayesian statistics, 
namely variational inference. Our proposed approach consists in treating
the network recurrent units as stochastic latent variables with a prior
distribution imposed over them. On this basis, we proceed to infer
corresponding posteriors; these can be used for prediction and recommendation
generation, in a way that accounts for the uncertainty in 
the available sparse training data. To allow for our approach
to easily scale to large real-world datasets, we perform inference under an approximate
amortized variational inference (AVI) setup, whereby the learned posteriors are parameterized via (conventional) neural networks. We perform an extensive experimental evaluation of our 
approach using challenging benchmark datasets, and illustrate its superiority over
existing state-of-the-art techniques.
\end{abstract}

%
%



\maketitle

\section{Introduction}

Recommender Systems (RS) constitute a key part of modern e-commerce websites \cite{session}; their aim is to enhance the user experience, by providing personalized product recommendations. Recent study on RS has mainly focused on matrix factorization (MF) methods and neighborhood search-type models. Such approaches work well in practice when a rich user profile can be built from the available data. Unfortunately, though, rich user profiles seldom are readily available to real-world systems. 

Session-based recommendation is a characteristic challenge that cannot be properly addressed by conventional methodologies employed in the context of RS. Specifically, under a session-based
setup, recommendation is based only on the actions of a user during a specific browsing session \cite{session}. Indeed, this type of recommendation generation approach is based on tracking user actions during an active session. Based on the captured and inferred session-based user behavioral patterns, it tries to predict the following user actions during that session, and proactively recommend items/actions to them.

From this description, it becomes apparent that session-based recommendation engines
attempt to generate effective recommendations with the availability of user-specific data being 
extremely limited.  Consequently, under this setting, conventional algorithmic approaches towards RS are confronted with hard challenges that stem from the unavailability of rich user profiles (data sparsity) \cite{koren}. Hence, in order to obtain effective session-based RS, it is imperative that novel methodological approaches be devised. Such methods must be capable of more effectively inferring and leveraging subtle session patterns, with the ultimate goal of enriching the available user profiles so as to properly address the challenges associated with data sparsity. 

Indeed, user session data constitute action sequences potentially entailing rich, complex, and subtle temporal dynamics. Thus, enabling effective extraction of these underlying dynamics, and utilizing them in the context of a preference inference mechanism, may result in novel session-based RS with considerably improved recommendation quality performance compared to the alternatives. Markov chain models (e.g., \cite{mdp}) constitute the most typical type of machine learning methods used to achieve this goal. However, recent breakthroughs in the field of Deep Learning (DL) \cite{lecun} have lately
come into close scrutiny, as a potential alternative means of addressing these challenges. Specifically, the introduction of novel treatments of Recurrent Neural Networks (RNNs) \cite{rnn-nade}, with compelling performance in as challenging and diverse tasks as image recognition, natural language understanding, and video captioning, has motivated their application to session-based RS. Contrary to simple deep network formulations that comprise only feedforward connections, RNNs also entail recurrent connections that allow for them to construct an internal representation of their observations history \cite{rnn}. This representation, which is encoded in the form of a high-dimensional vector of hidden unit activations, can then be utilized to address challenging learning tasks dealing with sequential data. 

In this context, the work recently presented in \cite{rnn-session} constitutes the most characteristic development. The RNNs employed therein are presented with data regarding the items
a user selects and clicks on during a given session. On this basis, recommendation 
relies on the history of previous actions (clicks on items) during that session, and the inferred behavioral
patterns. As shown therein, this method yields state-of-the-art session-based recommendation performance in several challenging benchmark problems. 

Motivated from these advances, in this paper we seek derivation of a solid inferential framework that 
allows for increasing the capability of RNN-driven session-based RS to ameliorate the
negative effects of data sparsity. To this end, we draw inspiration from recent RS developments which rely on the utilization of Bayesian inference techniques (e.g., \cite{acml12, cikm, cikm13, bpnmf, dpm-side}). Bayesian inference in the context of RS can be performed by considering that the postulated model variables pertaining to the system users and items are stochastic latent variables with some prior distribution imposed over them. This inferential framework allows for the developed recommendation
engine to account for the uncertainty in the available (sparse) training data. 
Thus, it is expected to allow for much improved predictive performance outcomes compared to the alternatives.

Under this rationale, our proposed approach is founded upon the fundamental assumption
that the hidden units of the postulated RNNs constitute latent variables of stochastic nature,
imposed some appropriate prior distribution. On this basis, we proceed to infer their corresponding
posteriors, using the available training data. Specifically, to allow for our model to scale to real-world
datasets, comprising millions of examples, we perform inference by resorting to the 
amortized variational inference (AVI) paradigm \cite{aevb, aevb2}. This is an approximate inference approach, which consists in: (i) parameterizing the inferred posterior distributions by means of conventional neural networks (inference networks); and (ii) casting the inference problem as an optimization problem, by making use of ideas from variational calculus. 

We evaluate the efficacy of the so-obtained approach, dubbed Recurrent 
Latent Variable Network for Session-Based Recommendation (ReLaVaR), 
considering a challenging publicly available benchmark.
We compare the obtained predictive performance of ReLaVaR with 
state-of-the-art techniques; we show that our approach completely outperforms
the competition, without presenting any limitations in terms of computational
efficiency and scalability.

The remainder of this paper is organized as follows: In the following Section, we 
provide a brief overview of the related work. In Section 3, we introduce our 
approach; specifically, we elaborate on its motivation, formally define our proposed model,
and derive its training and prediction generation formulae. In Section 4, we 
perform the experimental evaluation of our approach, and illustrate its merits 
over the current state-of-the-art. Finally, in the concluding Section of this paper,
we summarize our contribution and discuss our results.

\section{Related Work}

The continuous explosion in the availability of content through the
Internet renders design of RS a significant challenge 
for both the academic and industrial research communities. 
Ratings-based collaborative filtering (CF) systems have
served as an effective approach to address this challenge
 \cite{cf}. The intuitive idea behind their design consists in considering 
that the preferences of a user can be inferred by exploiting past
ratings of that user as well as users with related behavioral patterns.
This thriving subfield of machine learning started becoming popular
in the late 1990s with the development of online services such 
as Amazon, Yahoo! Music, MovieLens, Netflix, and CiteULike. 

Existing CF methods can be classified intro three main categories:
memory-based methods, model-based methods, and hybrid methods which
attempt to combine ideas from the former two paradigms. Memory-based
systems generate predictions under a neighborhood-driven rationale:
Item rating prediction for a target user comprises determination of
other users with similar ratings (target user neighbors), and computation of a weighted average
of the ratings of each item provided by the target user neighbors \cite{am1,am2}. A 
drawback of this paradigm is that, given the high sparsity of the
ratings matrix, the neighborhood of a target user may contain only
few, if any, ratings for a given item. Moreover, such approaches require
keeping the whole ratings matrix in memory to perform prediction in
real-time. This might be computationally prohibitive when dealing
with large real-world systems; thus, scalability is limited. 

Model-based CF methods attempt to ameliorate these issues by using
the available ratings data to train a machine learning model which expresses the
rating decision function of the users. Given the trained model,
prediction generation becomes extremely efficient, thus affording
scalable real-time operation. MF-based methods are perhaps the most popular 
class of model-based CF approaches \cite{decoste,vbmf,bpmf,bpnmf}.
These methods assume that the registered users and items are related
to sets of variables that lie in some low-dimensional latent space;
prediction is performed based on these latent features inferred for
each user and item. 

Recently, several authors have considered
introducing elaborate statistical assumptions into MF, that 
allow for performing full Bayesian inference 
(e.g., \cite{acml12, cikm, cikm13, bpnmf, dpm-side}). Under this approach,
it is considered that the user and item variables constitute stochastic latent
variables, over which an appropriate prior distribution is imposed,
and a corresponding posterior is inferred from the data.
Broad empirical evidence has shown that, under such a 
Bayesian inference-driven setup, real-world RS can yield
a noticeable predictive accuracy improvement without 
comprises in computational scalability. Indeed, this outcome
is well-expected from a theoretical point of view; this is due to the 
fact that a Bayesian inference treatment allows for better accounting for the uncertainty in the 
(training) data, which is prevalent in RS due to the 
sparsity of the available ratings matrices. 

On the other hand, in the last years the field of machine learning has witnessed a new 
wave of innovation, due to the DL breakthrough \cite{lecun}.
Unsurprisingly, the significant advances accomplished in the context of DL
have had a noteworthy impact on the ongoing research on RS. Indeed,
several researchers working on model-based CF methods have recently proposed
novel CF algorithms that employ DL-based models as an alternative to conventional
MF-driven approaches. 

In this vein, the work of \cite{rbm} constitutes one of the earliest ones 
that adopted ideas from the field of DL to effect the CF task. Specifically, they employed
Restricted Boltzmann Machines (RBMs) to learn the user and item latent vectors, and showed 
that their approach outperforms various popular alternatives in the Netflix challenge dataset.
More recently, the method in \cite{wang} presented a hierarchical Bayesian model called collaborative deep learning (CDL) for RS. This approach attempts to resolve the cold-start problem by 
augmenting the MF algorithm with appropriate side information related to item content.
This side information is obtained, in turn, from a DL model; this learns to 
extract useful, high-level representations from the raw item content, so as to inform the MF 
algorithm. 

Despite this extensive research effort devoted to  RS, session-based recommendation is a field that 
remained unappreciated for quite long, and has only recently attracted significant attention from the research community. Indeed, most of news and media sites, as well as many e-commerce sites (especially of small retailers) track the users that visit their sites only for short periods of time. Besides, the use of cookies or browser fingerprinting does not allow for obtaining 
reliable user data over long periods, spanning multiple sessions. Finally, it is very often the case that the behavior of users exhibits session-based traits. 

These facts bring to the fore the need of developing effective session-based RS, that 
can satisfy the following desiderata: (i) each session of the same user must be treated independently of their previous ones; (ii) the used algorithms must be capable of extracting subtle temporal behavioral patterns from the available user profiles, e.g. item-to-item similarity, co-occurrence, and transition probabilities; and (iii) this inferential procedure must be effectively carried out over long temporal horizons, as opposed to unrealistic low-order (e.g., one-step) temporal dependence models, that take only the last click or selection of the user into account (and ignore the information of past clicks in the same session).

To address these issues, \cite{zhang} introduced a novel framework based on traditional RNNs, and evaluated it using the click-through logs of a large scale commercial engine; their results showed significant improvements on the click-prediction accuracy compared to sequence-independent approaches. In a similar vein, \cite{rnn-session} presented an RNN-type machine learning model capable of learning subtle temporal patterns in user session data obtained from large e-commerce websites. Specifically, to allow for effectively extracting
high-order temporal dynamics, they utilized Gated Recurrent Unit
(GRU) networks \cite{gru}. Such networks entail a more elaborate model of an RNN unit, that aims at dealing with the vanishing/exploding gradient problem; this is a problem that plagues training of conventional RNNs, often rendering it completely infeasible \cite{lstm}. Their method was shown to outperform state-of-the-art alternatives in two large-scale tasks, including the challenging RecSys Challenge 2015 benchmark \cite{benchmark}. Finally, \cite{tan} proposed two extensions of the breakthrough work of \cite{rnn-session}, namely: (i) data augmentation via sequence preprocessing; and (ii) a simple model pre-training technique, to account for temporal shifts in the data distribution. As they showed, their proposed extensions yield an improvement over the method in \cite{rnn-session} by more than 10\%.

\section{Proposed Approach}
As we have already discussed, the main contribution of our work consists 
in devising a machine learning model that is capable of extracting subtle
and informative temporal dynamics from sparse user session data, with a special 
focus on accounting for the entailed uncertainty; we use this 
information to drive the developed recommendation algorithm.
We posit that such a capacity will allow for a significant increase in the
eventually obtained predictive performance.

Under this rationale, ReLaVaR frames the session-based recommendation problem
as a sequence-based prediction problem. Specifically, let us denote as 
$\{\boldsymbol{x}_{i}\}_{i=1}^{n}$ a user session; here, $\boldsymbol{x}_{i}$ is the
$i${th} clicked item, which constitutes a selection among $m$ alternatives,
and is encoded in the form of an 1-hot representation. Then, we formulate session-based recommendation as the problem of predicting the score vector $\boldsymbol{y}_{i+1}=[y_{i+1,j}]_{j=1}^{m}$ of the available items with respect to the following user action, where $y_{i+1,j} \in \mathbb{R}$ is the predicted score of the $j$th item. Typically, we are interested in recommending more than one items for the considered user to choose from; hence, at each time point we select the top-$k$ items (as ranked via $\boldsymbol{y})$ to recommend to the user. Thus, the goal of this work is to devise a machine learning model capable of more accurately predicting the vectors $\boldsymbol{y}_{i+1}$, given the observed subsequences $\{\boldsymbol{x}_{s}\}_{s=1}^{i}, \, \forall i$.

\subsection{Methodological Background}
To achieve our goals, we draw inspiration from state-of-the-art, RNN-based approaches,
such as \cite{rnn-session}. Thus, we begin by postulating an RNN structure comprising GRU units. At each time point, $i$, the postulated network is presented with the current user action (selected item), $\boldsymbol{x}_{i}$, and is expected to generate a prediction for the score vector $\boldsymbol{y}_{i+1}$ pertaining to the $(i+1)$th user selection. Formally, the recurrent units activation vector, $\boldsymbol{h}$, of a postulated GRU-based network are updated at time $i$ according to the following expression:
\begin{equation}
\boldsymbol{h}_{i} = (1-\boldsymbol{z}_{i}) \cdot \boldsymbol{h}_{i-1} + \boldsymbol{z}_{i} \cdot \hat{\boldsymbol{h}}_{i}
\end{equation}
where $\cdot$ denotes the elementwise product between two vectors, $\boldsymbol{h}_{i-1}$ is the recurrent units activation vector at the previous time point, and $\boldsymbol{z}_{i}$ is the \emph{update gate} output. This gate essentially learns to control when and by how much to update the hidden state of the recurrent units; it holds
\begin{equation}
\boldsymbol{z}_{i} = \tau(\boldsymbol{W}_{z} \boldsymbol{x}_{i} + \boldsymbol{U}_{z} \boldsymbol{h}_{i-1}) 
\end{equation}
where $\tau()$ is the logistic sigmoid function, and the $\boldsymbol{W}_{z}$ and $\boldsymbol{U}_{z}$ are trainable network parameters. On the other hand, in Eq. (1), $\hat{\boldsymbol{h}}_{i}$ is the \emph{candidate activations vector} of the GRU units at time $i$; its expression is a standard recurrent unit update expression with trainable parameters $\boldsymbol{W}$ and $\boldsymbol{U}$, yielding
\begin{equation}
\hat{\boldsymbol{h}}_{i} = \mathrm{tanh} (\boldsymbol{W} \boldsymbol{x}_{i} + \boldsymbol{U} (\boldsymbol{r}_{i} \cdot \boldsymbol{h}_{i-1})) 
\end{equation}
Here, $\boldsymbol{r}_{i}$ denotes the output of the \emph{reset gate} of the GRU network. This gate essentially learns to 
decide when the internal memory of the GRU units must be reset, with the ultimate goal of 
preventing the gradients of the model objective function from exploding to infinity or vanishing to zero
during model training; it reads
\begin{equation}
\boldsymbol{r}_{i} = \tau(\boldsymbol{W}_{r} \boldsymbol{x}_{i} + \boldsymbol{U}_{r} \boldsymbol{h}_{i-1}) 
\end{equation}
with the $\boldsymbol{W}_{r}$ and $\boldsymbol{U}_{r}$ being trainable network parameters.

\subsection{Model Formulation}

ReLaVaR extends upon the model design principles discussed in the previous section. It does so by introducing a novel formulation that renders the developed GRU-based recommendation model amenable to Bayesian inference. To effect our modeling goals, we consider that the component recurrent unit activations are stochastic latent variables. Specifically, we start by imposing a simple prior distribution over them, which reads
\begin{equation}
p(\boldsymbol{h}_{i}) = \mathcal{N}(\boldsymbol{h}_{i} | \boldsymbol{0}, \boldsymbol{I})
\end{equation}
where $\mathcal{N}(\boldsymbol{\xi} | \boldsymbol{\mu}, \boldsymbol{\Sigma})$ is a multivariate
Gaussian density with mean $\boldsymbol{\mu}$ and covariance matrix $\boldsymbol{\Sigma}$, 
and $\boldsymbol{I}$ is the identity matrix.

On this basis, we seek to devise an efficient means of inferring the corresponding posterior distributions, given the available training data. To this end, we draw inspiration from the AVI paradigm \cite{aevb}; specifically, we postulate that the sought posteriors, $q(\boldsymbol{h})$, approximately take the form of Gaussians with means and isotropic covariance matrices parameterized via GRU networks. We have:
\begin{equation}
q(\boldsymbol{h}_{i}; \boldsymbol{\theta}) = \mathcal{N}(\boldsymbol{h}_{i} | \boldsymbol{\mu_{\theta}}(\boldsymbol{x}_{i}),
\sigma_{\boldsymbol{\theta}}^{2}(\boldsymbol{x}_{i})\boldsymbol{I} )
\end{equation}

In this expression, the mean vectors, $\boldsymbol{\mu_{\theta}}(\boldsymbol{x}_{i})$, as well as the  variance functions, $\sigma_{\boldsymbol{\theta}}^{2}(\boldsymbol{x}_{i})$, are outputs of a postulated GRU network, with parameters set $\boldsymbol{\theta}$. In other words, we have
\begin{equation}
[\boldsymbol{\mu_{\theta}}(\boldsymbol{x}_{i}), \mathrm{log}\, \sigma_{\boldsymbol{\theta}}^{2}(\boldsymbol{x}_{i})] = (1-\boldsymbol{z}_{i}) \cdot [\boldsymbol{\mu_{\theta}}(\boldsymbol{x}_{i-1}), \mathrm{log}\, \sigma_{\boldsymbol{\theta}}^{2}(\boldsymbol{x}_{i-1})]  + \boldsymbol{z}_{i} \cdot \hat{\boldsymbol{h}}_{i}
\end{equation}
where
\begin{equation}
\boldsymbol{z}_{i} = \tau(\boldsymbol{W}_{z} \boldsymbol{x}_{i} + \boldsymbol{U}_{z} [\boldsymbol{\mu_{\theta}}(\boldsymbol{x}_{i-1}), \mathrm{log}\, \sigma_{\boldsymbol{\theta}}^{2}(\boldsymbol{x}_{i-1})]) 
\end{equation}
\begin{equation}
\hat{\boldsymbol{h}}_{i} = \mathrm{tanh} (\boldsymbol{W} \boldsymbol{x}_{i} + \boldsymbol{U} (\boldsymbol{r}_{i} \cdot [\boldsymbol{\mu_{\theta}}(\boldsymbol{x}_{i-1}), \mathrm{log}\, \sigma_{\boldsymbol{\theta}}^{2}(\boldsymbol{x}_{i-1})])) 
\end{equation}
and
\begin{equation}
\boldsymbol{r}_{i} = \tau(\boldsymbol{W}_{r} \boldsymbol{x}_{i} + \boldsymbol{U}_{r} [\boldsymbol{\mu_{\theta}}(\boldsymbol{x}_{i-1}), \mathrm{log}\, \sigma_{\boldsymbol{\theta}}^{2}(\boldsymbol{x}_{i-1})]) 
\end{equation}
while $[\boldsymbol{\xi}, \boldsymbol{\zeta}]$ denotes the concatenation of vectors 
$\boldsymbol{\xi}$ and $\boldsymbol{\zeta}$. 
On this basis, the values of the latent variables (stochastic unit activations) $\boldsymbol{h}_{i}$ 
can be computed by drawing (posterior) samples from the inferred posterior density (6).

Finally, let us turn to the output layer of the proposed model. 
This is presented with the (drawn samples of the) activation vectors 
$\boldsymbol{h}_{i}$ of our model, and generates
a vector of predicted score values $\boldsymbol{y}_{i+1}$ pertaining to the following
user action. On this basis, we need to appropriately impose a suitable distribution over
these generated output variables of our model, $\boldsymbol{y}_{i+1}$, conditional on the 
corresponding latent vectors, $\boldsymbol{h}_{i}$. Indeed, by reviewing the related literature, one may discover a number of possible alternatives for the conditional likelihood function of a ranking prediction model with the kind of probabilistic formulation that ReLaVaR adopts. Each one of these alternatives essentially gives rise to a different rationale in terms of quantifying the ranking accuracy of the trained model. 

In general, item ranking can be pointwise, pairwise or listwise. Pointwise ranking estimates the score of items independently of each other; then, the goal of model training is to ensure that relevant items receive a high score. Pairwise ranking compares the score of pairs of a positive and a negative item; then, model training aims at enforcing the score of the positive item to be higher than that of the negative one, for all the available pairs. Such a construction allows for one to limit score computation for the purposes of model training to a select subset of the available items. On the downside, such a formulation may undermine the eventually obtained accuracy of the recommendation algorithm. 
On the other hand, pointwise approaches require score computation for the whole set of available items. This is certainly more computationally demanding than pairwise approaches. However, this extra computational complexity does not necessarily translate into reduced scalability to real-world systems. This is especially the case with DL algorithms, which can be easily parallelized at a large scale by using cheap GPU hardware. Finally, listwise ranking uses the scores of all items and compares them to the perfect ordering. This entails item sorting, which can be computationally prohibitive in cases of large-scale systems.

Motivated from this discussion, in this work we resort to the most straightforward conditional likelihood selection for our model, namely a simple Multinoulli distribution; that is
\begin{equation}
p({y}_{i+1,j}=1 | \boldsymbol{h}_{i}) \propto \tau(\boldsymbol{w}_{y}^{j} \cdot \boldsymbol{h}_{i})
\end{equation}
where $\boldsymbol{W}_{y}=[\boldsymbol{w}_{y}^{j}]_{j=1}^{m}$ are trainable parameters of the output layer of the model. 

This selection can be viewed as giving rise to a pointwise ranking criterion, with 
the associated ranking loss function, $L_{s}$, being equal to the negative conditional log-likelihood expression that stems from (11). We have:
\begin{align}
L_{s} = & -\sum_{i=1}^{n}\mathrm{log}\,p(\boldsymbol{y}_{i+1}|\boldsymbol{h}_{i}) \\
- & \sum_{i,j=1}^{n,m} \big\{ y_{i+1,j} \; \mathrm{log} \, p({y}_{i+1,j}=1 | \boldsymbol{h}_{i}) \\
&- (1-y_{i+1,j}) \; \mathrm{log} \, (1-p({y}_{i+1,j}=1 | \boldsymbol{h}_{i})) \big\}
\end{align}
It is easy to notice that this loss function expression, $L_{s}$,  essentially constitutes the familiar (binary) Cross-Entropy function, widely used in DL literature.

\subsection{Training Algorithm}

Let us consider a training dataset $\mathcal{D}$, which comprises a number of 
click sequences (sessions), pertaining to a multitude of users. Variational inference for the 
developed ReLaVaR model consists in maximization of a lower-bound to the log-marginal likelihood of the model (evidence lower-bound, ELBO) w.r.t. the model parameters \cite{vbg}. 
Based on the previous model definition, the ELBO expression of ReLaVaR yields:
\begin{equation}
\mathrm{log}\,p(\mathcal{D})\geq\sum_{i}\bigg\{-\mathrm{KL}\big[q(\boldsymbol{h}_{i};\boldsymbol{\theta})||p(\boldsymbol{h}_{i})\big]-\mathbb{E}[L_{s}]\bigg\}
\end{equation}
Here, $\mathrm{KL}\big[q||p\big]$ is the KL divergence between the
distribution $q(\cdot)$ and the distribution $p(\cdot)$ [its analytical expression can be found in the Appendix].

Unfortunately, the posterior expectation $\mathbb{E}[L_{s}]$ cannot
be computed analytically; hence, its gradient becomes intractable. This is due to the nonconjugate formulation of ReLaVaR, which stems from its nonlinear assumptions. As a consequence, training the entailed parameter sets $\boldsymbol{\theta}$ is infeasible.

One could argue that this problem might be resolved by approximating this expectation using 
a set of $\Gamma$ Monte-Carlo (MC) samples, $\{\boldsymbol{h}_{i}^{\gamma}\}_{\gamma=1}^{\Gamma}$, drawn from the inferred posteriors. However, it is well-known that such an approximation would result in estimators with unacceptably high variance. AVI resolves these issues by means of a smart reparameterization of the MC samples of a Gaussian posterior density; specifically, we have \cite{aevb, aevb2}:
\begin{equation}
\boldsymbol{h}^{\gamma}=\boldsymbol{\mu}_{\boldsymbol{\theta}}(\cdot) +\sigma_{\boldsymbol{\theta}}(\cdot)\, \boldsymbol{\epsilon}^{\gamma}
\end{equation}
where $\boldsymbol{\epsilon}^{\gamma}$ is white random noise with
unitary variance, i.e. $\boldsymbol{\epsilon}^{\gamma}\sim\mathcal{N}(\boldsymbol{0},\boldsymbol{I})$.
By adopting this reparameterization, the MC samples drawn from the posterior density (6) can be now expressed as differentiable functions of the sought parameter sets, $\boldsymbol{\theta}$, and some random noise variable with low (unitary) variance, $\boldsymbol{\epsilon}$. Consequently, the problematic posterior expectation $\mathbb{E}[L_{s}]$, originally defined over the latent activations, reduces to a much more attractive
posterior expectation w.r.t. a low variance (random noise) variable. 

Then, by taking the gradient
of the so-reparameterized ELBO (13) in the context of any stochastic optimization algorithm, one can yield low variance estimators for the parameter sets, under some mild conditions \cite{aevb}. To this end, \cite{aevb} suggest utilization of Adagrad; this constitutes
a stochastic gradient algorithm with adaptive step-size \cite{adagrad}, and fast and proven convergence to a local optimum. We follow this advice in this work, selecting Adagrad as the stochastic optimizer of choice for training the ReLaVaR model.
  
\subsection{Prediction Generation}
Having trained a ReLaVaR model, given some dataset $\mathcal{D}$, recommendation generation
in the context of a user session can be performed by computing the predicted ratings $\boldsymbol{y}_{i+1} $, and selecting the top-$k$ of them to recommend to the user. To effect this procedure, we sample
the latent variables $\boldsymbol{h}_{i}$ from the corresponding variational posterior distributions. Indeed, to allow for obtaining 
reliable estimators, we draw a set comprising $\Gamma$ samples from the posteriors (6);
eventually, this yields a set of scoring function samples, $\{\boldsymbol{y}_{i+1}^{\gamma}\}_{\gamma=1}^{\Gamma}$. Then, recommendation is performed on the basis of the mean of these samples; that is, we employ a standard MC-type rationale.

\section{Experimental Evaluation}

To provide strong empirical evidence of the merits of our approach, in this Section we extensively evaluate it in challenging experimental scenarios. To this end, we exploit the benchmark dataset 
released in the context of the RecSys Challenge 2015 \cite{benchmark}; this comprises click-stream data pertaining to user sessions with an e-commerce website. 

Unfortunately, the test set of the aforementioned benchmark 
does not provide groundtruth information that can be used for recommendation quality evaluation.  
To resolve this issue, we adopt the solution suggested in \cite{rnn-session}; we split the originally available training data into one set comprising 7,966,257 sessions (with a total of 31,637,239 click actions), and another one comprising the remainder 5,324 sessions (with a total of 71, 222 click actions); we use the former for model training and the latter for evaluation purposes. Both sets entail a total of 37,483 items that a user may select to click on. Thus, we are dealing with a very sparse dataset, where the need of inferring more subtle and informative patterns comes to the fore with increased complexity. 

To obtain some comparative results, apart from our method we also cite the performance of recently proposed state-of-the-art alternatives in the same benchmark, namely the GRU-based method presented in \cite{rnn-session}, and the M2 and M4 approach introduced in \cite{tan}. In addition, we also run on the same dataset 
a standard baseline method in the field of matrix factorization, namely BPR-MF \cite{bpr} and capture its accuracy. Since BPR-MF is designed for processing single item feature vectors, as oppposed to \emph{sequences}, to apply it in the context of session-based recommendation we average the feature vectors of the items occurring thus far in a given session.

Our source codes have been developed in Python, and made use of the
Theano library\footnote{\url{http://deeplearning.net/software/theano/}}
\cite{Bastien-Theano-2012}. We run our experiments on an Intel Xeon
2.5GHz Quad-Core server with 64GB RAM and an NVIDIA Tesla K40 GPU
accelerator. 

\subsection{ReLaVaR Model Configuration}
In the following, we experiment with a diverse set of selections for the size of the latent variable space (number of recurrent latent variables). We report the best performing model configuration in Section 4.3; we provide deep insights on how model performance varies with this selection in Section 4.4. In all cases, parameter initialization for our model is performed by resorting to the Glorot Normal initialization scheme \cite{glorot}; dropout with a rate equal to 0.5 is employed for regularization purposes. 

To perform model training, Adagrad is carried out by utilizing \emph{session-parallel mini-batches}, following the suggestions in \cite{rnn-session}. Let us consider we adopt a mini-batch size equal to $\beta$. Then, session-parallel mini-batches can be obtained by using the first event of the first $\beta$ sessions to form the input data of the first mini-batch (the desired output is the second events of our active sessions); we use the second events to form the second mini-batch, and so forth. When a session ends, we put the next available session in its place. In the occasion of such a switch taking place, we reset the appropriate hidden state of the model, since we assume that the training sessions constitute independent and identically distributed (sequential) data. 

To facilitate convergence, Nesterov momentum \cite{momentum} is additionally applied during training. In all cases, Adagrad's global stepsize is chosen from the set $\{0.005,0.01,0.05,0.1\}$, while momentum strength is chosen from the set $\{0,0.1,0.2,0.3,0.4\}$, both on the basis of network performance on the training set in the first few training algorithm iterations. Contrary to \cite{tan}, we do not perform any tedious data augmentation procedure or model pretraining. Thus, our approach is not \emph{directly comparable} to \cite{tan}, since application of the pre-processing steps proposed therein should be well-expected to also increase the performance of ReLaVaR. However, we do cite the performance results reported in \cite{tan} for completeness sake. 

\subsection{Performance Metrics}
To quantitatively assess the performance of our approach, we employ two
commonly used evaluation metrics, namely Recall@20 and Mean Reciprocal Rank (MRR)@20. 
The former metric expresses the frequency at which the desired (groundtruth) item 
in the test data makes it to the 20 highest ranked items suggested by the evaluated approach.
Hence, this metric allows for modeling and assessing certain practical scenarios where there is no highlighting of recommendations; what matters is the desired item being included in a short list of recommendations, rather than the absolute order that these items are presented to the user.
On the other hand, MRR@20 describes the average predicted score of the desired items 
in the test data, with the score values set to zero if the desired item does not make
it to the top-20 list of ranked items. Thus, MRR@20 models scenarios where absolute item
ordering does matter; for instance, it allows for better algorithm evaluation in cases where  
the lower ranked items are visible only after scrolling.

\subsection{Empirical Performance}
We commence the presentation of our experimental results by reporting on the 
best-performing configuration of our model (i.e., selection of the number of latent variables that maximizes empirical performance on the test set); our findings
are summarized in Table 1. In the same Table, we also illustrate how these
empirical findings compare to the considered competitors of our method, that is BPR-MF, M2, M4,
and the GRU-driven approach of \cite{rnn-session}. We report two different performance 
results for the GRU-based method, which correspond to two different loss functions considered 
in \cite{rnn-session}, namely BPR and TOP1; the former selection yields 
better Recall@20 outcomes for that method, while the latter yields a better MRR@20 value. Moreover, regarding the methods proposed in \cite{tan} it can be observed from Table 1 that as the number of hidden units increases the accuracy of the proposed method declines. 

As we observe, our approach outperforms all previously reported state-of-the-art results in terms of the 
MRR@20 metric, while yielding the second-best reported performance in terms of the Recall@20 metric\footnote{We obtained this outcome with the mini-batch size set equal to 50, Adagrad step size set equal to 0.05, and momentum strength set equal to 0.}. The number of component latent variables for our method is equal to 1500; this represents a larger network than the one that obtains best performance for the considered alternatives. This fact constitutes further supporting evidence of the capacity of our approach to extract subtler temporal dynamics from the available data without getting prone to overfitting.

\begin{table}

\caption{Best performance results of the evaluated methods.}

\begin{centering}
\begin{tabular}{|c|c|c|c|}
\hline 
Method & Model Size  & Recall@20 & MRR@20 \tabularnewline
\hline 
\hline
BPR-MF  & - & 0.2574 & 0.0618 \tabularnewline
\hline 
GRU w/ BPR Loss \cite{rnn-session}& $ 1000$ & 0.6322 & 0.2467 \tabularnewline
\hline 
GRU w/ TOP1 Loss \cite{rnn-session}& $ 1000$ & 0.6206 & 0.2693 \tabularnewline
\hline 
M2 \cite{tan}& $ 100$ & 0.7129 & 0.3091 \tabularnewline
\hline 
M4 \cite{tan}& $ 1000$ & 0.6676 & 0.2847 \tabularnewline
\hline 
ReLaVaR & $ 1500$  & 0.6507 & 0.3527\tabularnewline
\hline 
\end{tabular}
\par\end{centering}
\end{table}

\subsection{Further Investigation}

\subsubsection{Varying the size of the latent variable space}
It is well-understood that the number of latent units bears significant impact on 
the obtained empirical performance. To allow for examining the extent of 
this effect, in Fig. 1 we show how the considered performance
metrics vary when adjusting the number of latent units of a trained ReLaVaR model. 
These results have been obtained with the training algorithm hyperparameters remaining the same as described in section 4.3.

As it can be seen from Fig. 1, ReLaVaR accuracy, as measured via both the considered metrics, grows as we add more latent units. The increase is more prominent when the network size is small, and tends to be lower for larger networks. 

\begin{figure}
\begin{centering}
\includegraphics[scale=0.55]{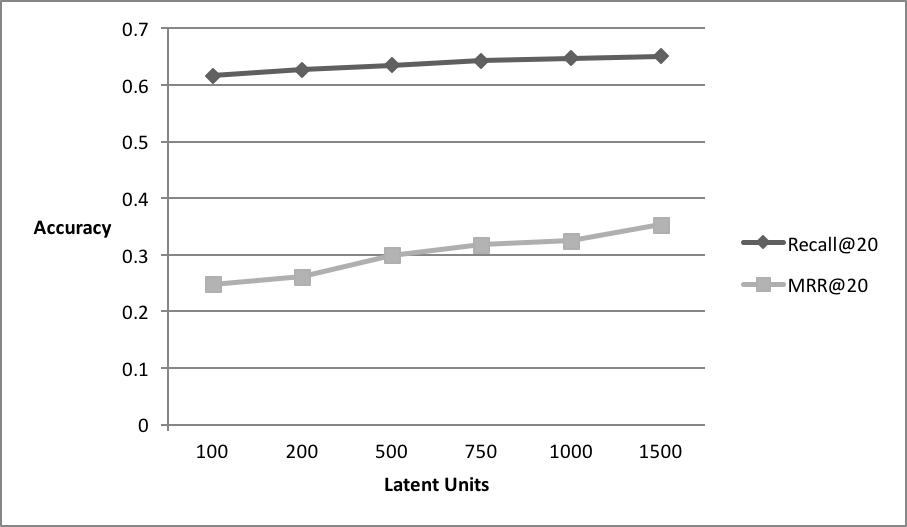}
\par\end{centering}
\caption{ReLaVaR performance fluctuation with the number of latent
variables.}

\end{figure}

\subsubsection{Considering alternative loss functions}
Further, it is interesting to examine how the performance of our model compares to the competition
in case we adopt a different type of loss function, $L_{s}$. To this end, we consider a pairwise ranking loss, namely the TOP1 loss function that was introduced in \cite{rnn-session}.

The obtained results (for best model configuration) are provided in Table 2. As we observe, 
appropriate selection of the employed loss function, $L_{s}$, is a crucial factor that 
determines the success of our approach in modeling the considered dataset. Indeed,
replacement of the Cross-Entropy loss function with TOP1 yields a notable performance deterioration, especially in terms of the obtained MRR@20 values. 

To provide some further insights, in Fig. 2 we illustrate the corresponding results regarding performance fluctuation with the size of the latent variable space. We observe that model size continues to have a significant effect on ReLaVaR predictive performance when using this alternative loss function.

\begin{table}

\caption{ReLaVaR model performance for different selections of the employed loss function, $L_{s}$ (results correspond to best network configuration).}

\begin{centering}
\begin{tabular}{|c|c|c|c|}
\hline 
Loss Function  & TOP1 & Cross-Entropy\tabularnewline
\hline 
\hline 
\# Latent Units & 1000 & 1500\tabularnewline
\hline 
Step Size & 0.1 & 0.05\tabularnewline
\hline 
Momentum Weight & 0 & 0\tabularnewline
\hline 
\hline 
Recall@20 & 0.6250 & 0.6507\tabularnewline
\hline 
MRR@20 & 0.2727 & 0.3527\tabularnewline
\hline 
\end{tabular}
\par\end{centering}
\end{table}

\begin{figure}
\begin{centering}
\includegraphics[scale=0.55]{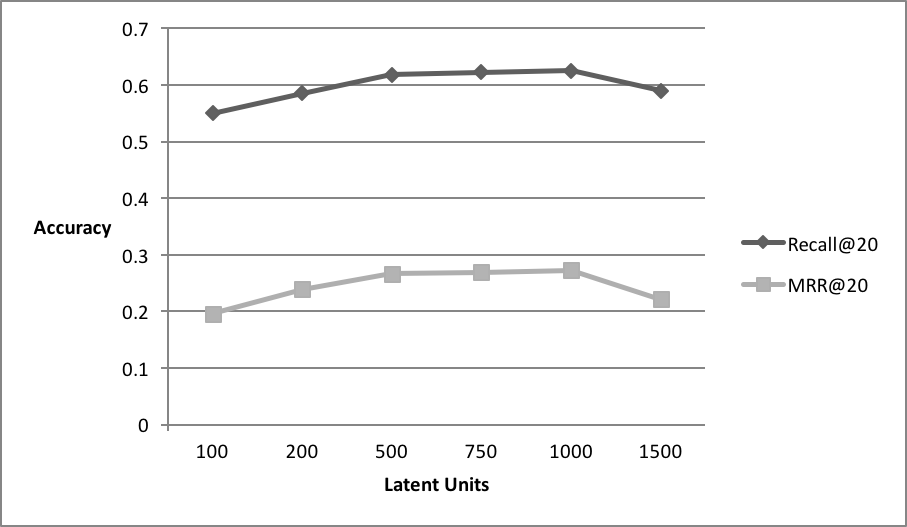}
\par\end{centering}
\caption{ReLaVaR performance fluctuation with the number of latent
variables: Use of the TOP1 loss function.}

\end{figure}

\subsection{Computational Complexity}
Apart from predictive accuracy, another aspect of machine learning models that is of 
utmost importance when dealing with real-world applications concerns computational efficiency. 
This aspect entails examination of both model scalability to large datasets,
as well as of the imposed computational overheads when it comes to prediction generation.

To investigate these aspects of the proposed approach, in Table 3 we perform a comparison of wall-clock times between our method and the current state-of-the-art. The measurements reported therein concern alternative selections of the employed loss function, $L_{s}$; in all cases, they pertain to network sizes yielding the best empirical accuracy of the evaluated methods. 

From this exhibition, it becomes apparent that not only our approach yields competitive accuracy, but it does so while remaining competitive in terms of computational costs. Specifically, notice that the slight difference in computational costs between the original formulation of ReLaVaR (i.e., using the Cross-Entropy loss function) and the competition is merely due to the larger size of the trained network. 
Note also that all the compared approaches allow for real-time prediction generation. Thus, we can soundly argue that our method constitutes an attractive solution for building real-world session-based RS.

\begin{table*}
\caption{Comparison of computational times (in seconds), for various selections of the employed loss function. Results pertain to network configurations yielding best accuracy.}

\begin{centering}
\begin{tabular}{|c|c|c|c|}
\hline 
 Method & Network Size & Training time (total) & Prediction time per click event (average) \tabularnewline
\hline 
\hline 
 GRU w/ BPR Loss & 1000 Units & 48692.48 & 0.683 \tabularnewline
\hline 
 GRU w/ TOP1 Loss & 1000 Units & 44716.60 & 0.627 \tabularnewline
\hline 
\hline 
ReLaVaR w/ TOP1 Loss  &  1000 Units & 42357.84 & 0.595 \tabularnewline
\hline 
ReLaVaR w/ Cross-Entropy Loss & 1500 Units & 60109.86 & 0.844 \tabularnewline
\hline 
\end{tabular}
\par\end{centering}
\end{table*}

\section{Conclusions and Future Work}
In this paper, we attacked the problem of session-based recommendation. Specifically, our work was
motivated from the sequential nature of the addressed predictive setup, and the associated 
sparsity of the available data. Our expectation was that, by better addressing these issues, one may be able to obtain a noticeable improvement in the quality of the generated recommendations.

To this end, we introduced a way of improving the modeling capacity of RS that utilize deep learning techniques with recurrently connected units. Specifically, we effected this goal by adopting concepts from the field of Bayesian statistics, namely variational inference. The proposed approach, dubbed ReLaVaR, constitutes a hierarchical latent variable model, where the inferred posterior distributions are parameterized via GRU networks. Such a Bayesian inferential setup: (i) retains the prowess of  
existing GRU-based networks in terms of extracting and analyzing salient temporal patterns 
in the available user sessions data; and (ii) allows for accounting for the uncertainty in the available (sparse) data when performing prediction and recommendation generation; enabling this capability is well-known to yield a noticeable performance improvement in real-world data modeling scenarios. 

We performed an extensive experimental evaluation of our approach using a challenging benchmark dataset. We provided thorough insights into the predictive behavior of our approach under different setups of the employed training algorithm, as well as under different selections of the postulated network configuration. As we showed, our approach is capable of outperforming existing state-of-the-art alternatives in terms of two popular performance metrics. We also illustrated that our proposed approach achieves this accuracy improvement without undermining computational efficiency, both in training time and in prediction generation time.

One research direction that we have not considered in this work concerns the possibility of stacking multiple network layers, one on top of the other, to create a more potent sequential data modeling pipeline. In such a formulation, the input of the bottom GRU  layer is the observed data sequence, $\{\boldsymbol{x}_{i}\}_{i=1}^{n}$; on the other hand, each one of the subsequent GRU layers is presented with the sequence of activation vectors, $\{\boldsymbol{h}_{i}\}_{i=1}^{n}$, of the layer that immediately precedes it in the hierarchy; the model output layer is driven from the recurrent unit activations vector of the topmost GRU layer. 

Stacking multiple layers of GRU networks allows for performing inference and analysis of temporal patterns in multiple time-scales. Thus, one might theoretically expect that such an architecture should be capable of yielding improved performance in real-world session-based RS. However, related findings in the recent literature, e.g. \cite{rnn-session}, have shown this not to be the case; for instance, \cite{rnn-session} showed that (stacked) multilayer architectures impose significant extra computational burden without yielding any benefit in terms of predictive accuracy. Indeed, this behavior can be attributed to the short typical length of user sessions, whence temporal pattern analysis on multiple time-scales becomes less relevant. 

It was these results that motivated us not to examine stacked multilayer variants of ReLaVaR in the context of this work. However, we do believe that stacked multilayer variants of ReLaVaR might possess favorable performance characteristics in the context of different sorts of systems dealing with session data. Thus, investigation of such possible opportunities remains among our plans for future research pursuits.

\section*{Appendix}
Using (5)-(6), we obtain:
\begin{equation}
\begin{split}
\mathrm{KL}\big[q(\boldsymbol{h}_{i};\boldsymbol{\theta})||p(\boldsymbol{h}_{i})\big] = &
-\frac{1}{2}\sum_{d=1}^{D} {\left[ \boldsymbol{\mu_{\theta}}(\boldsymbol{x}_{i})^{2}\right]_{d}} \\ &+
\frac{D}{2} \big[1+ \mathrm{log}\,\sigma_{\theta}(\boldsymbol{x}_{i})^{2} -\sigma_{\boldsymbol{\theta}}(\boldsymbol{x}_{i}) ^{2}\big]
\end{split}
\end{equation}
where $[\cdot]_{d}$ is the $d$th element of a vector, and $D$ is the dimensionality of the postulated latent space.

\section*{Acknowledgment}
We gratefully acknowledge the support of NVIDIA Corporation with the
donation of one Tesla K40 GPU used for this research.

\bibliographystyle{ACM-Reference-Format}
\bibliography{CF} 

\end{document}